\RequirePackage{etoolbox}
\csdef{input@path}{%
 {sty/}
 {pics/}
 {sections/}
}%

\documentclass[ba]{imsart}
\pubyear{2018}
\volume{00}
\issue{0}
\doi{0000}
\firstpage{1}
\lastpage{28}

\usepackage{amsthm}
\usepackage{amsmath}
\usepackage{natbib}
\usepackage[colorlinks,citecolor=blue,urlcolor=blue,filecolor=blue,backref=page]{hyperref}
\usepackage{graphicx}

\usepackage{url}
\makeatletter
\g@addto@macro{\UrlBreaks}{\UrlOrds}

\RequirePackage{accents}
\RequirePackage{amsfonts}

\newtheorem{prop}{Proposition}

\usepackage[usenames,dvipsnames,svgnames,table]{xcolor}
\usepackage[T1]{fontenc}
\usepackage{booktabs}
\usepackage{nth}
\usepackage{subfig}
\usepackage{placeins}
\usepackage{cite}
\usepackage{float}

\startlocaldefs
\numberwithin{equation}{section}
\theoremstyle{plain}

\endlocaldefs
\allowdisplaybreaks
\usepackage{bbm}

\begin{document}

\begin{frontmatter}
\title{Bayesian method for causal inference in spatially-correlated multivariate time series}
\runtitle{Bayesian multivariate time series causal inference}

\begin{aug}
\author{\fnms{Bo} \snm{Ning}\thanksref{addr1}\ead[label=e1]{bning@ncsu.edu}},
\author{\fnms{Subhashis} \snm{Ghosal}\thanksref{addr1}\ead[label=e2]{sghosal@ncsu.edu}}
\and
\author{\fnms{Jewell} \snm{Thomas}\thanksref{addr2}\thanksref{addr3}
\ead[label=e3]{jewellt@live.unc.edu}}

\runauthor{Bo Ning et al.}

\address[addr1]{Department of Statistics,
North Carolina State University,
2501 Founders Drive,
Raleigh, North Carolina 27695,
USA.
\printead{e1},
\printead{e2}}

\address[addr2]{Department of English \& Comparative Literature,
The University of North Carolina at Chapel Hill,
CB \#3520 Greenlaw Hall,
Chapel Hill, North Carolina 27599,
USA.
\printead{e3}}

\address[addr3]{Jewell Thomas is a former Staff Data Scientist at MaxPoint Interactive Inc., 3020 Carrington Mill Blvd, Morrisville, North Carolina 27560, USA.}

\end{aug}

\begin{abstract}

Measuring the causal impact of an advertising campaign on sales is an essential task for advertising companies.
Challenges arise when companies run advertising campaigns in multiple stores which are spatially correlated, and when the sales data have a low signal-to-noise ratio which makes the advertising effects hard to detect.
This paper proposes a solution to address both of these challenges.
A novel Bayesian method is proposed to detect weaker impacts and
a multivariate structural time series model is used to capture the spatial correlation between stores through placing a $\mathcal{G}$-Wishart prior on the precision matrix.
The new method is to compare two posterior distributions of a latent variable---one obtained by using the observed data from the test stores and the other one obtained by using the data from their counterfactual potential outcomes.
The counterfactual potential outcomes are estimated from the data of synthetic controls,
each of which is a linear combination of sales figures at many control stores over the causal period. 
Control stores are selected using a revised Expectation-Maximization variable selection (EMVS) method.
A two-stage algorithm is proposed to estimate the parameters of the model.
To prevent the prediction intervals from being explosive,
a stationarity constraint is imposed on the local linear trend of the model through a recently proposed prior. 
The benefit of using this prior is discussed in this paper.
A detailed simulation study shows the effectiveness of using our proposed method to detect weaker causal impact.
The new method is applied to measure the causal effect of an advertising campaign for a consumer product sold at stores of a large national retail chain. 

\end{abstract}



\end{frontmatter}

\section{Introduction}
\label{sec:intro}

Advertising is thought to impact sales in markets.
{\it MaxPoint Interactive Inc.} ({\it MaxPoint}),
an online advertising company,\footnote{The methodology developed and 
presented in this paper is not connected to any commercial 
products currently sold by {\it MaxPoint}.} 
is interested in measuring the sales increases associated with running advertising campaigns for products distributed through
brick-and-mortar retail stores. 

\begin{figure}[!h]
\centering
   \includegraphics[width=6cm]{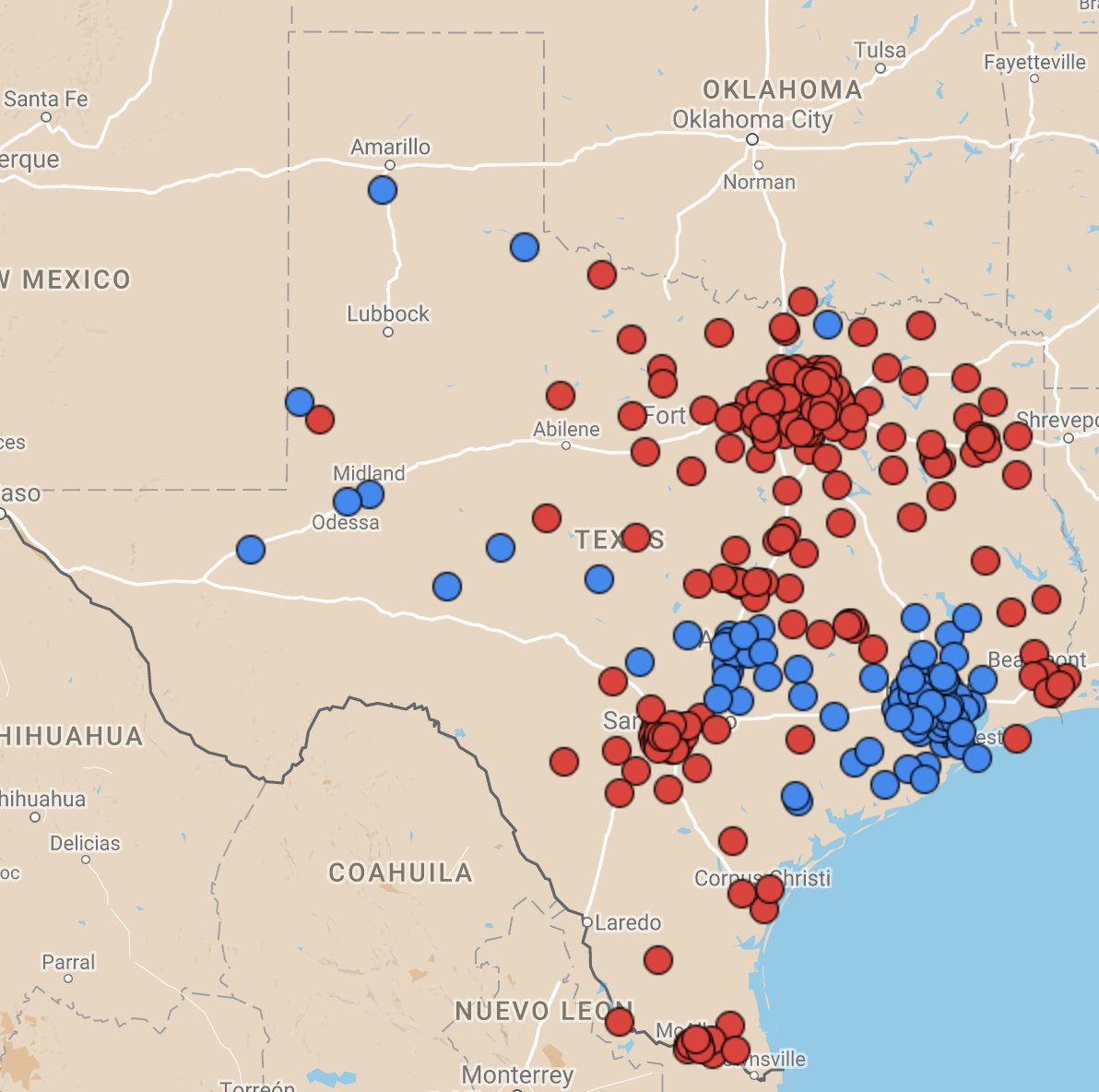}
   \caption{\small An example of test and control store locations in the State of Texas (Google Maps, 2017).
   The red dots represent the locations of the test stores;
   the blue dots represent the locations of the control stores. }
   \label{fig-0}
\end{figure}

The dataset provided by {\it Maxpoint} was obtained as follows: 
{\it MaxPoint} ran an advertising campaign at 627 test stores across the United States. 
An additional 318 stores were chosen as control stores.
Control stores were not targeted in the advertising campaign.
The company collected weekly sales data from all of these stores for 36 weeks before
the campaign began and for the 10 weeks in which the campaign was conducted.
The time during which the campaign was conducted
is known.
The test stores and the control stores were randomly selected from different economic regions across the U.S..
Figure \ref{fig-0} shows an example of the locations of stores in the State of Texas.\footnote{Note: The locations of the
stores shown in the figure are not associated with any real datasets collected by {\it MaxPoint}.}

To the best of our knowledge,
the work of \citet{brod15} is the most related one to the present study. 
Their method can be described as follows. 
For each test store,
they first split its time series data into two parts: before and during a causal impact (in our case, the impact is the advertising campaign).
Then, they used the data collected before the impact to predict the values during the causal period. At the same time, they applied a stochastic search variable selection (SSVS)
method to construct a synthetic control for that store.
The counterfactual potential outcomes \citep{rubi05} are the sum of 
the predicted values and the data from the synthetic control.
Clearly, the potential outcomes of the store exposed to advertising were the observed data.
Finally, they compared the difference
between the two potential outcomes and took the average of differences across different time points. 
The averaged difference is a commonly used causal estimand
that measures the temporal average treatment effect \citep{boji17}.

The method proposed by \citet{brod15} is novel and attractive; however, it cannot directly apply to analyze our dataset
due to the following three reasons: 
(1) Many causal impacts in our dataset are weak.
The causal estimand that \citet{brod15} used often fails to detect them;
(2) The test stores within an economic region are spatially correlated as they share similar demographic information.
Using \citet{brod15}'s method would not allow to consider the spatial correlation between stores;
(3) The SSVS method is computationally slow because it
requires sampling from a large model space consisting of $2^p$ possible combinations of $p$ control stores.
In the following, we will discuss our proposed method for addressing these three difficulties.

First, we propose a new method for detecting weaker causal impacts.
The method compares two posterior distributions of the latent variables of the model, where one distribution is computed by conditioning the observed data and the other one is computed by conditioning the counterfactual potential outcomes.
We use the one-sided Kolmogorov-Smirnov (KS) distance to quantify the distance between the two posterior distributions.

The new method can successfully detect weaker impacts because it compares two potential outcomes at the latent variable level; while the commonly used method compares them at the observation level.
Since the observed data often contain ``inconvenient'' components---such as seasonality and random errors---which 
inflate the uncertainty of the estimated causal effect, 
the commonly used method may fail to detect weaker impacts.
In the simulation study, we show that the new method outperforms the commonly used method even when the model is slightly misspecified. 

The causal estimand in the new method is different from the causal estimand of the commonly used method.
The former one measures the temporal average treatment effect using the KS distance between two posterior distributions
and
the latter measures that effect using the difference between two potential outcomes.
Formal definitions of the two causal estimands are provided in Section \ref{sec-1}.

Secondly, 
we use a multivariate version of a structural time series model \citep{harv90} to model the sales data of test stores by allowing pooling of information among those stores that locate in geographically contiguous economic regions.
This model enjoys a few advantages that make it especially suitable for our causal inference framework.
First, the model is flexible to adapt to different structures of the latent process.
Secondly, it can be written as a linear Gaussian state-space model and exact posterior sampling methods can be carried out by applying the Kalman filter and simulation smoother algorithm proposed by \citet{durb02, durb01}.
Thirdly, it is relatively easy to deal with missing data due to the use of the Kalman filter and backward smoothing (KFBS) algorithm. 
The imputing process can be naturally incorporated into the 
Markov chain Monte Carlo (MCMC) loop.

Since test stores are correlated, the number of parameters in the covariance matrix grows quadratically with the dimension. 
Consequently, there will not be enough data to estimate all these parameters.
In our approach, we reduce the number of parameters by imposing sparsity based on a given spatial structure \citep{smit07, barb15, li15}.
We consider a graphical model structure for dependence based on geographical distances  
between stores. If the distance between two stores is very large, we treat them conditionally independent given other stores. In terms of a graphical model, this is equivalent to not put an edge between them.
We denote the corresponding graph by $\mathcal{G}$. Note that $\mathcal{G}$ is given in our setting and is completely determined by the chosen thresholding procedure. We use a graphical Wishart prior with respect to the given graph $\mathcal{G}$, in short a $\mathcal{G}$-Wishart prior \citep{rove02}, to  impose sparsity on the precision matrix. 
One advantage is that this prior is conjugate for a multivariate normal distribution.
If $\mathcal{G}$ is decomposable, sampling from a conjugate $\mathcal{G}$-Wishart posterior is relatively easy due to an available closed form expression for the normalizing 
constant in the density \citep{laur96, rove00, rove02}. 
However, if $\mathcal{G}$ is non-decomposable, 
the normalizing constant does not usually have a simple closed form 
\citep[see however;][]{uhle17}, and thus one cannot easily sample directly
from its posterior. 
In such a situation, an approximation for the normalizing constant is
commonly used 
\citep{atay05, mits11, wang12, khar15}.
A recent method introduced by
\citet{Moha15} is a birth-death Markov chain Monte Carlo (BDMCMC) 
sampling method.
It uses a trans-dimension MCMC algorithm that transforms sampling
of a high-dimensional matrix to lower dimensional matrices, thus improving 
efficiency when working with large precision matrices.

In a multivariate state-space model, 
the time dynamics are described by a multivariate stochastic trend, 
usually an order-one vector autoregressive (VAR(1)) process
\citep{jong91, jong94, koop97, durb02}. 
To use a VARMA($p, q$) 
(order $p$ vector autoregression with order $q$ moving average) 
with $p > 1$, $q \geq 0$ process is also possible and the choice of 
$p, q$ can be made based on data (e.g., chosen by the Bayesian Information Criterion). 
However, the larger the $p$ and $q$ are, the larger the number of parameters that need to be estimated.
For the sake of tractability, we treat the hidden process as a VAR(1) process throughout the paper.

Putting stationarity constraints on the VAR(1) process is necessary to prevent the prediction intervals from becoming too wide to be useful.
However, constructing an appropriate prior complying with the constraints is not straightforward.
\citet{gelf92} proposed a
naive approach that puts a conjugate prior on the vector autoregressive parameter
to generate samples
and only keep the samples meeting the constraints.
However, it can be highly inefficient when many draws from the posterior 
correspond to nonstationary processes. 
A simple remedy is to project these nonstationary draws  
on the stationarity region to force them to meet the constraints \citep{gunn05}. 
However, the projection method is somewhat unappealing 
from a Bayesian point of view because it would 
make the majority of the projected draws
have eigenvalues lying on the boundary of the corresponding space \citep{gali06, roy16}. 
We instead follow the recently proposed method of \citet{roy16} to decompose
the matrix into several unrestricted parameters 
so that commonly used priors can be put on 
those parameters. 
While conjugacy will no longer be possible, 
efficient algorithms for drawing samples from the posterior distribution are available. 

Thirdly, to accelerate the computational speed of selection control stores, 
we suggest using a revised version of the Expectation-Maximization variable selection (EMVS) method \citep{rock14}.
The model uses an Expectation-Maximization (EM) algorithm that is faster and does not need to search $2^p$ possible combinations.

It is worth mentioning that there are many other popular methods for constructing a synthetic control,
such as the synthetic control method proposed by \citet{abad03},
the difference-in-differences method \citep{abad05, bonh11, dona07},
and the matching method \citep{stua10}. 
Moreover, \citet{doud16} provided a nice discussion on the advantages and disadvantages of each method.
Unlike
these methods, there are two advantages of using our proposed method:
It does not need to have a prior knowledge about the relevant control stores,
the process of selecting control stores is completely driven by data
and can be easily incorporated into a Bayesian framework.
It provides a natural model-based causal inference by viewing counterfactual potential outcomes as missing values and generating predicting credible intervals from 
their posterior predictive distributions, and finally providing a quantitative measure for the strength of the causal effect
\citep{rubi05}.

We apply our method on both simulated datasets and the real dataset provided by {\it MaxPoint}.
In the simulation study, 
we compare the new method with the method proposed by \citet{brod15}.

The rest of the paper is organized as follows. 
Section \ref{sec-1} introduces causal assumptions and causal estimands.
Section \ref{sec-2} describes the model and the priors. 
Section \ref{sec-3} describes posterior computation techniques. 
Section \ref{sec-4} introduces our proposed new approach to infer causal effects in times series models.
Simulation studies are conducted in Section \ref{sec-5}. In 
Section \ref{sec-6}, the proposed method is applied on a real dataset from 
an advertising campaign conducted by {\it MaxPoint}. 
Finally, Section \ref{sec-7} concludes with a discussion.

\section{Causal assumptions and causal estimands}
\label{sec-1}

This section includes three parts.
First, we will introduce the potential outcomes framework.
Secondly, we shall discuss three causal assumptions. 
Finally, we shall define two causal estimands, one of them is new.

The potential outcomes framework is widely used in causal inference literature \citep{rubi74, rubi05, ding17}.
Potential outcomes are defined as the values of an outcome variable at a future point in time after treatment under two different treatment levels.
Clearly, at most one of the potential outcomes for each unit can be observed, and the rest are missing 
\citep{holl86, rubi78, imbe15}.
The missing values can be predicted using statistical methods.
In the paper, we predict the values using the data from a synthetic control
that is constructed from several control stores.

Based on the potential outcomes framework, we conduct the causal inference. 
There are three assumptions need to make for conducting the inference. They are,
\begin{enumerate}
    \item The stable unit treatment value assumption (SUTVA);
    \item The strong ignorability assumption on the assignment mechanism;
    \item The trend stays stable in the absence of treatment for each test store.
\end{enumerate}

The SUTVA contains two sub-assumptions: 
no interference between units and 
no different versions of a treatment \citep{rubi74}.
The first assumption is reasonable because the stores did not interact with each other after the advertising was assigned.
As \citet{paul07} pointed out, ``interference is distinct from statistical dependence produced by pretreatment clustering.''
Since the spatial correlation between test stores is produced by pretreatment clustering, it is different from the interference between stores.
The second assumption is also sensible because we assume
that there are no multiple versions of the advertising campaign.
For example, the advertising campaign is not launched across multiple channels.

The strong ignorability assumption also contains two parts: 
unconfoundedness and positivity \citep{ding17}.
Unconfoundedness means that the treatment is assigned randomly 
and positivity means that the probability for each store being assigned is positive.
In our study, we assume the company randomly assigned advertising to stores and each store has an equal probability of being assigned.

The last assumption says that the counterfactual potential outcomes in the absence of the advertising in test stores are predictable. 

Now, we shall introduce some notations before defining causal estimands. 
Let $n$ be the total number of test stores to which the advertising were assigned.
The $i$-th test store has $p_i$ control stores (stores did not assigned with the advertising), $i = 1, \dots, n$. 
The total number of control stores are denoted as $p$,
$p = \sum_{i=1}^n p_i$.
The length of the time series data is $T+P$. 
Let $1, \dots, T$ be the periods before running the advertising campaign and
$T+1, \dots, T+P$ be the periods during the campaign. 
Let $\boldsymbol w_t = (w_{1t}, \dots, w_{n+p,t})'$ be a vector of treatment
at time $t = T+1, \dots, T+P$, 
with each $w_{it}$ being a binary variable.
The treatment assignment is time-invariant,
so $\boldsymbol w_t = \boldsymbol w$.
For stores assigned with advertising, 
we denote the sales value
for the $i$-th store at times $t$ as $y_{it}$.
Let $y_{it}^{\text{obs}}$ be the observed data and
$y_{it}^{\text{cf}}$ be the counterfactual potential outcomes which are missing.
We let 
$\boldsymbol Y_t^{\text{obs}} = 
(y^{\text{obs}}_{1t}, \dots, y^{\text{obs}}_{nt})'$
and 
$\boldsymbol Y_t^{\text{cf}} = 
(y^{\text{cf}}_{1t}, \dots, y^{\text{cf}}_{nt})'$
respectively be the observed and
missing potential outcomes for $n$ test stores at time $t$,
$t = 1, \dots, T+P$.
Clearly,
$\boldsymbol Y_t^{\text{obs}}
= 
\boldsymbol Y_t^{\text{cf}}$
when $t = 1, \dots, T$.
We define
$\boldsymbol Y_{T+1: T+P}^{\text{obs}} = 
(\boldsymbol Y^{\text{obs}}_{T+1}, \dots, 
\boldsymbol Y^{\text{obs}}_{T+P})'$
and 
$\boldsymbol Y_{T+1: T+P}^{\text{cf}} = 
(\boldsymbol Y^{\text{cf}}_{T+1}, \dots, 
\boldsymbol Y^{\text{cf}}_{T+P})'$.

We first define the causal estimand of a commonly used method.
For the $i$-th test store, the commonly used causal estimand 
is defined as 
\begin{equation*}
    \frac{1}{P}\sum_{t = T+1}^{T+P}
    \big(y_{it}^{\text{obs}} - y_{it}^{\text{cf}}\big),
\end{equation*}
which is the temporal average treatment effects \citep{boji17} 
at $P$ time points.
In our setting, the treatment effects for $n$ test stores are defined as
\begin{equation}
\label{causal-estimand-1}
    \frac{1}{P}\sum_{t = T+1}^{T+P}
    \big(\boldsymbol Y_t^{\text{obs}} - \boldsymbol Y_t^{\text{cf}}\big).
\end{equation}

To introduce our new causal estimand,
let $x_{it}$ be the data for the synthetic
control for the $i$-th test store at time $t$. 
Recall that the data of a synthetic control is a weighted sum of the sales from several control stores.
Define 
$\boldsymbol X_{1:T+P} = (\boldsymbol X_1, \dots, \boldsymbol X_{T+P})$,
where $\boldsymbol X_t$ is an $n \times p$ matrix
containing data from $p$ control stores at time $t$.
Let $\mu_{it}$ be a latent variable of a model, which is of interest.
Define 
$\boldsymbol \mu_{t} = (\boldsymbol \mu_{1t}, 
\dots, \boldsymbol \mu_{nt})$
which is an $n \times 1$ vector.
We let 
\begin{equation}
\label{causal-estimand-2-1}
    p(\sum_{t=T+1}^{T+P} \boldsymbol \mu_t
    \big| 
\boldsymbol Y_{1:T+P}^{\text{obs}},
\boldsymbol X_{1:T+P}).
\end{equation}
be the posterior distribution of the latent variable conditional on 
$\boldsymbol Y_{1:T+P}^{\text{obs}}$
and $\boldsymbol X_{1:T+P}$,
and
\begin{equation}
\label{causal-estimand-2-2}
 p(\sum_{t=T+1}^{T+P} \boldsymbol \mu_t
    \big| 
\boldsymbol Y_{1:T}^{\text{obs}},
\boldsymbol Y_{T+1:T+P}^{\text{cf}},
\boldsymbol X_{1:T+P}).
\end{equation}
be the distribution conditional on $\boldsymbol Y_{1:T+P}^{\text{cf}}$
and $\boldsymbol X_{1:T+P}$.

The new causal estimand is defined 
as the one-sided Kolmogorov-Smirnov (KS)
distance between the two distributions for $i$-th store, which can be expressed as
\begin{equation*}
\begin{split}
	& \sup_x \Big[  
	\mathcal{F}(\sum_{t=T+1}^{T+P} 
	\mu_{it} \leq x
	\big | 
	y_{i,1:T}^{\text{obs}},
	y_{i,T+1:T+P}^{\text{cf}},
	x_{i,1:T+P}) \\
	& \quad \quad - 
	\mathcal{F}(\sum_{t=T+1}^{T+P} 
	\mu_{it} \leq x
	\big | 
	y_{i,1:T+P}^{\text{obs}},
	x_{i,1:T+P})
	\Big],
\end{split}	
\end{equation*}
where $\mathcal{F}(\cdot)$ stands for the corresponding cumulative distribution function.
In our setting, since test stores are spatially correlated, the causal effect of the $i$-th test store is defined as
\begin{equation}
\label{causal-estimand-2}
\begin{split}
	& \sup_x \Big[  
	\mathcal{F}(\sum_{t=T+1}^{T+P} 
	\mu_{it} \leq x
	\big | 
	\boldsymbol Y_{1:T}^{\text{obs}},
	\boldsymbol Y_{T+1:T+P}^{\text{cf}},
	\boldsymbol X_{1:T+P})\\
	& \quad \quad - 
	\mathcal{F}(\sum_{t=T+1}^{T+P} 
	\mu_{it} \leq x
	\big | 
	\boldsymbol Y_{1:T+P}^{\text{obs}},
	\boldsymbol X_{1:T+P})
	\Big].
\end{split}	
\end{equation}

A larger value of the one-sided KS distance implies a potentially larger 
scale of causal impact.
An impact is declared to be significant if the one-sided KS distance is larger than its corresponding threshold. 
The threshold is calculated based on several datasets that are randomly drawn from the posterior predictive distribution of (\ref{causal-estimand-2-2})
(See Section \ref{sec-4} for more details.)

We would like to mention that although the proposed method is applied to a multivariate time series model in this paper,
even in the context of a univariate model, the idea of comparing posterior distributions of latent variables appears to be new.
Generally speaking,
this idea can be adopted into many other applications with different Bayesian models as long as these models are described in terms of latent variables.

\section{Model and prior}
\label{sec-2}

\subsection{Model}

We consider a multivariate structural time series model given by (to simplify the notation,
we use $\boldsymbol Y_t$ instead of $\boldsymbol Y_t^{\text{obs}}$
in the current and the following sections),
\begin{equation}
\label{equ-2.1}
      \boldsymbol {Y}_t \ = \ \boldsymbol \mu_t + \boldsymbol \delta_t
      								  + \boldsymbol X_t \boldsymbol \beta
                                  + \boldsymbol \epsilon_t,
\end{equation}
where
$\boldsymbol Y_t$, $\boldsymbol \mu_t$, $\boldsymbol \delta_t$ and
$\boldsymbol \epsilon_t$ are $n \times 1$ vectors standing for
the response variable, trend, seasonality and measurement error respectively.
$n$ is the number of test stores,
$\boldsymbol X_t$ is an $n \times p$ matrix containing data from $p$ control stores at time $t$ and $\boldsymbol \beta$  is a sparse $p \times 1$ vector of regression coefficients, where $p$ can be very large. 
We allow each response in $\boldsymbol Y_t$ to have different number of 
control stores, 
and write
\begin{equation*}
\boldsymbol X_t 
=
\begin{pmatrix}
    x_{11, t} & \cdots & x_{1p_1,t} & 0 & \cdots & 0 & \cdots & 0 & \cdots & 0\\
    0 & \cdots & 0 & x_{21, t} & \cdots & x_{2p_2, t} & \cdots  & 0 & \cdots & 0\\
       & \ddots &  & & \ddots & & & & \ddots & \\ 
    0 & \cdots & 0 & 0 & \cdots & 0 & \cdots  & x_{n1, t} & \cdots & x_{np_n, t},
\end{pmatrix},
\end{equation*}
with $\sum_{i=1}^n p_i = p$.
Let $\boldsymbol \gamma = (\gamma_1, \dots, \gamma_p)$ be the 
indicator variable such that
$\gamma_j = 1$ if and only if $\beta_j \neq 0$. 
$\boldsymbol \epsilon_t$ is an independent and identically distributed (i.i.d) error process.

The trend of the time series is modeled as
\begin{equation}
\label{equ-2.2}
      \boldsymbol \mu_{t+1} \ = \ \boldsymbol \mu_t 
                                             + \boldsymbol \tau_t
                                             + \boldsymbol u_t,
\end{equation}                                             
where  $\boldsymbol \tau_t$ is viewed as a term replacing the slope of the linear trend at time $t$ to allow for a general trend, and $\boldsymbol u_t$ is an i.i.d. error process. 
The process $\boldsymbol \tau_t$ can be modeled as a stationary VAR(1) process, driven by the equation 
\begin{equation}
\label{equ-2.3}
       \boldsymbol \tau_{t+1} \ = \ \boldsymbol D + 
                \boldsymbol \Phi (\boldsymbol \tau_t - \boldsymbol D) +
                \boldsymbol v_t,
\end{equation}
where $\boldsymbol D$ is an $n \times 1$ vector and 
$\boldsymbol \Phi$ is an $n\times n$ matrix of the coefficients of the VAR(1) process with   
eigenvalues having modulus less than 1.
If no stationarity restriction is imposed on  $\boldsymbol \tau_t$, we model it by                                      
\begin{equation}
\label{equ-2.4}
      \boldsymbol \tau_{t+1} \ = \ \boldsymbol \tau_t 
                                              + \boldsymbol v_t,
\end{equation}
where $\boldsymbol v_t$ is an i.i.d. error process.

The seasonal component $\boldsymbol \delta_t$ in (\ref{equ-2.1}) 
is assumed to follow the evolution equation
\begin{equation}
\label{equ-2.5}
      \boldsymbol \delta_{t+1} \ = \ - \sum_{j=0}^{S-2} \boldsymbol \delta_{t-j}
                                                 + \boldsymbol w_t,
\end{equation}
where $S$ is the total length of a cycle and $\boldsymbol w_t$ is an i.i.d. error process. 
For example, for an annual dataset, 
$S = 12$ represents the monthly effect while 
$S = 4$ represents the quarterly effect.
This equation ensures that the summation of $S$ time periods of each variable
has expectation zero.

We assume that the residuals of (\ref{equ-2.1})--(\ref{equ-2.5}) 
are mutually independent and time invariant,
and are distributed as multivariate normals with mean $\boldsymbol 0_{n\times1}$ 
and covariance matrices
$\boldsymbol \Sigma$, 
$\boldsymbol \Sigma_u$, 
$\boldsymbol \Sigma_v$ 
and $\boldsymbol \Sigma_w$ respectively.

By denoting parameters
$\boldsymbol \alpha_t = (\boldsymbol \mu_t', \boldsymbol \tau_t',
\boldsymbol \delta_t', \cdots, \boldsymbol \delta_{t-S+2}')'$ and
$\boldsymbol \eta_t = (\boldsymbol u_t', \boldsymbol v_t', \boldsymbol w_t')'$,
the model can be represented as a linear Gaussian state-space model
\begin{eqnarray}
\label{equ-2.6}
   \boldsymbol Y_t & = &  \boldsymbol z \boldsymbol \alpha_t 
   									+ \boldsymbol X_t \boldsymbol \beta
                                 + \boldsymbol \epsilon_t, \\
\label{equ-2.7}                                 
   \boldsymbol \alpha_{t+1} & = & \boldsymbol c
                                                + \boldsymbol T \boldsymbol \alpha_t 
                                                + \boldsymbol R\boldsymbol \eta_{t},                             
\end{eqnarray}
where $\boldsymbol z$, $\boldsymbol c$, 
$\boldsymbol T$ and $\boldsymbol R$ can be
rearranged accordingly based on the model (\ref{equ-2.1})--(\ref{equ-2.5});
and $\boldsymbol \epsilon_t \sim 
\mathcal{N}(\boldsymbol 0, \boldsymbol \Sigma)$,
$\boldsymbol \eta \sim \mathcal{N}(
\boldsymbol 0, \boldsymbol Q)$,
$\boldsymbol Q = \text{bdiag}(\boldsymbol \Sigma_u, \boldsymbol \Sigma_v, \boldsymbol \Sigma_w)$
are mutually independent; here and below ``$\mathrm{bdiag}$'' 
refers to a block-diagonal matrix with entries as specified.
If $\boldsymbol \tau_t$ is a nonstationary process in (\ref{equ-2.3}), 
then we set $\boldsymbol c = \boldsymbol 0$.

\subsection{Prior}
\label{sec-2.2}

We will now discuss the priors for the parameters in the model.
We separate the parameters into four blocks:
the time varying parameter $\boldsymbol \alpha_t$,
the stationarity constraint parameters $\boldsymbol D$ and $\boldsymbol \Phi$,
the covariance matrices of the error terms $\boldsymbol \Sigma$,
$\boldsymbol \Sigma_u$, $\boldsymbol \Sigma_v$ and $\boldsymbol \Sigma_w$,
and the sparse regression parameter $\boldsymbol \beta$.

For the time varying parameter, we give a prior 
$\boldsymbol \alpha_1 \sim \mathcal{N}(\boldsymbol a, \boldsymbol{P})$
with $\boldsymbol{a}$ is the mean and $\boldsymbol{P}$ is the covariance matrix.
For the covariance matrices of the errors, we choose priors as follows:
\begin{eqnarray*}
\boldsymbol \Sigma^{-1} \sim W_\mathcal{G}(\nu, \boldsymbol H),\ \ \ 
& \boldsymbol \Sigma_u^{-1}  \sim  W_\mathcal{G}(\nu, k_1^2(n+1)\boldsymbol H), \\
\boldsymbol \Sigma_v^{-1}  \sim  W_\mathcal{G}(\nu, k_2^2(n+1)\boldsymbol H),\ \ \ 
& \boldsymbol \Sigma_w^{-1}  \sim  W_\mathcal{G}(\nu, k_3^2(n+1)\boldsymbol H), 
\end{eqnarray*}
where $W_\mathcal{G}$ stands for a $\mathcal{G}$-Wishart distribution. 
For the stationarity constraint parameter $\boldsymbol D$,
we choose a conjugate prior
$\boldsymbol D 
\sim
\mathcal{N}(\boldsymbol 0, \boldsymbol I_n)$.

Putting a prior on the stationarity constraint matrix of a univariate AR(1) process is
straightfoward.
However, for the VAR(1) process in (\ref{equ-2.3}),
the stationarity matrix $\boldsymbol \Phi$ has to meet the Schur-stability constraint 
\citep{roy16}, that is, it needs to satisfy 
$|\lambda_j(\boldsymbol{\Phi})|<1$, $j=1,\ldots,n$, 
where $\lambda_j$ stands for the $j$th eigenvalue. 
Thus the parameter space of $\boldsymbol\Phi$ is given by 
\begin{equation}
\label{equ-2.8}
 \mathfrak{S}^n
 = 
 \{\boldsymbol \Phi \in 
 \mathbb{R}^{n \times n}: |\lambda_j(\boldsymbol \Phi)| < 1, j = 1, \dots, n \}. 
\end{equation}
Clearly simply putting a conjugate matrix-normal prior on $\boldsymbol \Phi$
does not guarantee that all the sample draws are Schur-stable.
We follow \citet{roy16}'s method of putting priors on 
$\boldsymbol \Phi$ through a representation as given below. 

We first denote
$\widetilde{\boldsymbol \tau}_t = \boldsymbol \tau_t - \boldsymbol D$,
then the Yule-Walker equation for $\widetilde{\boldsymbol \tau}_t$ is
\begin{equation}
\label{equ-2.9}
  \boldsymbol{U} = \boldsymbol \Phi \boldsymbol{U} \boldsymbol \Phi'
              + \boldsymbol \Sigma_v,
\end{equation}
where 
$\boldsymbol{U} = \mathbb{E}(\widetilde{\boldsymbol \tau}_t \widetilde{\boldsymbol \tau}_t')$ 
is a symmetric matrix.
Letting 
$f(\boldsymbol \Phi, \boldsymbol{U}) = \boldsymbol{U} - 
         \boldsymbol \Phi \boldsymbol{U} \boldsymbol \Phi'$,
we have that $f(\boldsymbol{\Phi}, \boldsymbol{U})$
is a positive definite matrix if and only if
$\boldsymbol \Phi \in \mathfrak{S}^n$ \citep{stei52}.  
Furthermore, we have the following proposition:

\begin{prop}
\label{prop-1}
[\citet{roy16}] Given a positive definite matrix $\boldsymbol{M}$, 
there exists a positive matrix $\boldsymbol{U}$,
and a square matrix $\boldsymbol \Phi \in \mathfrak{S}^n$ such that
$f(\boldsymbol \Phi, \boldsymbol{U}) = 
\boldsymbol{M}$ if and only if $\boldsymbol{U} \geq 
\boldsymbol{M}$ and
$\boldsymbol \Phi = (\boldsymbol{U} - \boldsymbol{M})^{1/2} \
\boldsymbol{O} \boldsymbol{U}^{-1/2}$ for an orthogonal matrix 
$\boldsymbol{O}$ with rank 
$r = \mathrm{rank}(\boldsymbol{U} - \boldsymbol{M})$,
where
$(\boldsymbol{U} - \boldsymbol{M})^{1/2}$ and 
$\boldsymbol{U}^{-1/2}$ are full column rank square root of matrices 
$(\boldsymbol{U} - \boldsymbol{M})$ and $\boldsymbol{U}^{-1}$.
\end{prop}

In view of Proposition \ref{prop-1},
given $\boldsymbol \Phi \in \mathfrak{S}^n$ and 
an arbitrary value of $\boldsymbol M$, 
the solution for $\boldsymbol{U}$ in equation (\ref{equ-2.9})
is given by
\begin{equation}
\label{equ-2.10}
  \text{vec}(\boldsymbol{U}) \ = \ (\boldsymbol I_{n^2} 
                     - \boldsymbol \Phi \otimes \boldsymbol \Phi)^{-1} 
                     \text{vec}(\boldsymbol{M}).
\end{equation}
Letting 
$\boldsymbol{V} = \boldsymbol{U} - \boldsymbol{M}$, 
we have $\boldsymbol \Phi 
= 
\boldsymbol{V}^{1/2} \boldsymbol{O} \boldsymbol{U}^{-1/2}$,
where $\boldsymbol{V}$ is a positive definite matrix, 
and $\boldsymbol{O}$ is an orthogonal matrix.
The matrix 
$\boldsymbol{V}$ can be represented by the 
Cholesky decomposition 
$\boldsymbol{V} 
= 
\boldsymbol{L} \boldsymbol{\Lambda}\boldsymbol{L}'$,
where $\boldsymbol{L}$ is a lower triangular matrix and 
$\boldsymbol\Lambda$ is a diagonal matrix with positive entries. 
Thus the number of unknown parameters in $\boldsymbol V$ 
reduces to ${n(n-1)}/{2} + n$.
The parameter $\boldsymbol{O}$ can be decomposed by using the Cayley representation  
\begin{equation}
\label{equ-2.11}
   \boldsymbol{O} 
   \ = \ 
   \boldsymbol{E}_\iota \cdot 
   [(\boldsymbol I_n - \boldsymbol{G})(\boldsymbol I_n + \boldsymbol{G})^{-1}]^2
\end{equation}
with 
$\boldsymbol{E}_\iota
= 
\boldsymbol I_n - 2\iota \boldsymbol{e}_1 \boldsymbol{e}_1'$, 
$\iota \in \{0, 1\}$ and $\boldsymbol{e}_1 = (1, 0, \dots, 0)'$, 
where $\boldsymbol{G}$ is a skew-symmetric matrix.
Thus the number of parameters in 
$\boldsymbol{O}$ is ${n(n-1)}/{2} + 1$. 
By taking the log-transform, the parameters in 
$\boldsymbol{\Lambda}$ can be made free of restrictions.
Therefore there are $n^2$ unrestricted parameters in $\boldsymbol \Phi$ plus one binary parameter.
We put normal priors on the $n^2$ unrestricted parameters: 
the lower triangular elements of $\boldsymbol L$,
the log-transformed diagonal elements of $\boldsymbol \Lambda$ and
the lower triangular elements of $\boldsymbol G$.
For convenience, we choose the same normal prior for those parameters and
choose a binomial prior for the binary parameter $\iota$.

For the sparse regression parameter $\boldsymbol \beta$,
we chose a spike-and-slab prior with
$\boldsymbol \beta \sim \mathcal{N}(\boldsymbol 0, \boldsymbol A_{\boldsymbol \gamma})$,
$\boldsymbol A_{\boldsymbol \gamma} = \text{diag}(a_1, \dots, a_p)$ with 
$a_i =  v_0 (1-\gamma_i) + v_1 \gamma_i$, 
where $0 \leq v_0  < v_1$, diag refers to a diagonal matrix with entries as specified; 
$\pi(\boldsymbol \gamma| \theta)
=
\theta^{|\boldsymbol \gamma|} (1-\theta)^{p-|\boldsymbol \gamma|}$
with $|\boldsymbol \gamma| = \sum_{i=1}^p \gamma_i$;
$\theta \sim \text{Beta}(\zeta_1, \zeta_2)$.

\section{Posterior computation}
\label{sec-3}

We propose a two-stage estimation algorithm to estimate the parameters.
In the first stage, we adopt a fast variable selection method
to obtain a point estimator for $\boldsymbol \beta$.
In the second stage, we plug-in its estimated value and
sample the remaining parameters using an MCMC algorithm.

To conduct the variable selection on $\boldsymbol \beta$,
a popular choice would be using a SSVS method \citep{geor93}.
The algorithm searches for $2^p$ possible combinations of
$\beta_i$ in $\boldsymbol \beta$ using Gibbs sampling 
under $\gamma = 0$ and $\gamma = 1$, $i = 1, \dots, p$. 
In the multivariate setting, this method is computationally very challenging when $p$ is large.
An alternative way is to use the EMVS method \citep{rock14}. 
This method uses the EM algorithm to maximize the posterior of 
$\boldsymbol \beta$
and thus obtain the estimated model.
It is computationally much faster than the SSVS method. Although SSVS gives a fully Bayesian method quantifying the uncertainty of variable selection through posterior distributions, the approach is not scalable for our application which involves a large sized data. Since quantifying uncertainty of variable selection is not an essential goal, as variable selection is only an auxiliary tool here to aid inference, the faster EMVS algorithm seems to be a pragmatic method to use in our application. 

After obtaining $\hat{\boldsymbol \beta}$,
we plug it into (\ref{equ-2.6})--(\ref{equ-2.7})
and deduct $\boldsymbol X_t \hat{\boldsymbol \beta}$ from $\boldsymbol Y_t$.
We denote the new data as $\widetilde{\boldsymbol Y}_t$,
and will work with the following model:
\begin{equation}
\label{equ-2.6new}
\begin{split}
	\widetilde{\boldsymbol Y}_t 
	& = 
	\boldsymbol z \boldsymbol \alpha_t 
		+ \boldsymbol \epsilon_t, \\
	\boldsymbol \alpha_{t+1} 
	& = \boldsymbol c + \boldsymbol T \boldsymbol \alpha_t
		+ \boldsymbol R \boldsymbol \eta_t.
\end{split}
\end{equation}
In the MCMC step, 
we sample the parameters in the Model 
(\ref{equ-2.6new}) from their corresponding posteriors.
Those parameters include:
the time-varying parameters $\boldsymbol \alpha_{1:T}$,
the stationarity constraint parameters $\boldsymbol D$ 
and $\boldsymbol \Phi$,
the covariance matrices of the residuals 
$\boldsymbol \Sigma^{-1}$, 
$\boldsymbol \Sigma^{-1}_u$,
$\boldsymbol \Sigma^{-1}_v$, and
$\boldsymbol \Sigma^{-1}_w$.

The details of the algorithm are presented in the 
supplementary material \citep{ning18}.

The proposed two-stage estimation algorithm is thus summarized as follows:
\begin{description}
\item {\bf Stage 1: EMVS step.}
Choose initial values for $\boldsymbol \beta^{(0)}$,
$\boldsymbol a_1^{*(0)}$ and 
$\boldsymbol P_1^{*(0)}$ 
using the revised EMVS algorithm to find the optimized value for $\boldsymbol \beta$.

\item 
{\bf Stage 2: MCMC step.}
Given $\widetilde{\boldsymbol Y}_t$, 
we sample parameters using MCMC with the following steps:
\begin{itemize}
\item[(a)] Generate $\boldsymbol \alpha_t$ using the Kalman filter and simulation smoother method. 

\item[(b)] Generate $\boldsymbol \Phi$ using the Metropolis-Hastings algorithm.

\item[(c)] Generate $\boldsymbol D$.

\item[(d)] Generate covariance matrices from their respective 
$\mathcal{G}$-Wishart posterior densities.

\item[(e)] Go to Step (a) and repeat until the chain converges.
\end{itemize}
\item 
Skip Step (b) and (c) if no stationarity restriction is imposed on
$\boldsymbol \tau_t$.
\end{description}

\section{A new method to infer causality}
\label{sec-4}

In this section, we will introduce our new method to infer causality (in short, ``the new method'') along a commonly used method.

Recall the treatment effects of the commonly used method is defined in (\ref{causal-estimand-1}).
Since $\sum_{t=T+1}^{T+P} \boldsymbol Y^{\text{cf}}_t$ is an unobserved quantity, 
we replace it by its posterior samples from 
$p(\sum_{t=T+1}^{T+P} \boldsymbol Y^{\text{cf}}_t | 
\boldsymbol Y_{1:T}^{\text{obs}}, \boldsymbol X_{1:T+P})$.

The commonly used method may fail to detect even for a moderately sized impact for two reasons. 
First, the prediction intervals increase linearly as the time lag increases.
Secondly, the trends are the only latent variables would give a response to an impact, 
including the random noise and the seasonality components would inflate the uncertainty of the estimated effect.
For the data have a low signal-to-noise ratio,
this method is which even harder to detect causal impacts.

We thus propose a new method by comparing only the posterior distributions of the latent trend in the model given the observations and the data from  counterfactuals.
The new method consists the following five steps:

 {\bf Step 1:} Applying the two-stage algorithm to obtain posterior samples
for parameters in the model using the data from the period without causal impacts.

{\bf Step 2:} Based on those posterior samples, obtaining 
sample draws of $\boldsymbol Y_{T+1:T+P}^{\text{cf}}$ 
from its predictive posterior distribution 
$p(\boldsymbol Y_{T+1:T+P}^{\text{cf}}| \boldsymbol Y_{1:T}^{\text{obs}},
\boldsymbol X_{1:T+P}$).

{\bf Step 3:} Generating $k$ different datasets from counterfactual potential outcomes (in short, ``counterfactual datasets'') from 
the predictive posterior distribution, for the $j$-th dataset, $j \in \{1, \dots, k\}$,
denoted by $\boldsymbol Y_{T+1:T+P}^{\text{cf}(j)}$.
Then fitting each $\boldsymbol Y_{T+1:T+P}^{\text{cf}(j)}$ 
into the model to obtain sample draws of the trend from
its posterior distribution,
which is shown in (\ref{causal-estimand-2-2})
(here, we replace $\boldsymbol Y_{T+1:T+P}^{\text{cf}}$ with
$\boldsymbol Y_{T+1:T+P}^{\text{cf}(j)}$).
Also, fitting the observed data 
$\boldsymbol Y_{1:T+P}^{\text{obs}}$
into the model and sampling from
(\ref{causal-estimand-2-1}).

{\bf Step 4:} Using the one-sided Kolmogorov-Smirnov (KS) distance to quantify the 
difference between the posterior distributions of the trend given by
the observed data and the counterfactual datasets. 
The posterior distribution of the trend given by the counterfactual datasets 
is obtained by stacking the sample draws estimated from all the $k$
simulated datasets.
Then calculating the KS distance
between the two posterior distributions 
for each store 
as follows:
\begin{equation}
\label{equ-4.1.2}
\begin{split}
	& \sup_x \Big[  
	\frac{1}{k} \sum_{j=1}^{k} \big(
	\mathcal{F}(\sum_{t=T+1}^{T+P} 
	\mu_{it} \leq x
	\big | 
	\boldsymbol Y_{1:T}^{\text{obs}},
	\boldsymbol Y_{T+1:T+P}^{\text{cf}(j)},
	\boldsymbol X_{1:T+P}) \big)\\
	& \quad \quad \quad \quad - 
	\mathcal{F}(\sum_{t=T+1}^{T+P} 
	\mu_{it} \leq x
	\big | 
	\boldsymbol Y_{1:T+P}^{\text{obs}},
	\boldsymbol X_{1:T+P})
	\Big],
\end{split}	
\end{equation}
where 
$i = 1, \dots, n$, and
$\mathcal{F}(\cdot)$ stands for the empirical distribution function of the obtained MCMC samples.

{\bf Step 5:} Calculating the $k \times (k-1)$ pairwise one-sided KS distances between 
the posterior distributions of the trends given by the $k$
simulated counterfactual datasets, 
that is to calculate the following expression
\begin{equation}
\label{equ-4.1.3}
\begin{split}
	& \sup_x \Big[ 
	\mathcal{F}(\sum_{t=T+1}^{T+P} 
	\mu_{it} \leq x
	\big | 
	\boldsymbol Y_{1:T}^{\text{obs}},
	\boldsymbol Y_{T+1:T+P}^{\text{cf}(j)},
	\boldsymbol X_{1:T+P}) \\
	& \quad  - 
	\mathcal{F}(\sum_{t=T+1}^{T+P} 
	\mu_{it} \leq x
	\big | 
	\boldsymbol Y_{1:T}^{\text{obs}},
	\boldsymbol Y_{T+1:T+P}^{\text{cf}(j')},
	\boldsymbol X_{1:T+P})
	\Big], 
\end{split}	
\end{equation}
where $j,j' = 1, \dots, k$, $j \neq j'$.
Then, for each $i$,
choosing the 95\% upper percentile among those distances as a threshold
to decide whether the KS distance calculated from (\ref{equ-4.1.2}) is significant or not.
If the KS distance is smaller than this threshold, 
then the corresponding causal impact is declared not significant.

The use of a threshold is necessary,
since the two posterior distributions of the trend obtained under
observed data and the data from the counterfactual are not exactly equal even when
there is no causal impact. Our method automatically selects a data-driven threshold through a limited repeated sampling as in multiple imputations.



So far we described the commonly used method and the new method in the setting where the period without a causal impact comes before that with the impact. 
However, the new method can be extended to allow datasets in more general situations when:
1) there are missing data from the period without causal impact;
2) the period without causal impact comes after the period with a impact;
3) there are more than one periods without causal impact,
 both before and after the period with a impact.
This is because the KFBS method is flexible to impute missing values at any positions in a times series dataset.

\FloatBarrier

\section{Simulation study}
\label{sec-5}

In this section, we conduct a simulation study to
compare the two different methods introduced in the last section. 
To keep the analysis simple, we only consider the setting that the period with causal impact follows that without the impact.
We also conduct convergence diagnostics for MCMC chains and a sensitivity analysis for the new method, the results are shown in Section 4 of the supplementary material \citep{ning18}.

\subsection{Data generation and Bayesian estimation}
\label{sec-4.1}
We simulate five spatially 
correlated datasets,
and assume the precision matrices in the model
have the adjacency matrix as follows:
\begin{equation}
\label{equ-4.1}
\begin{pmatrix}
1 & 1 & 0 & 0 & 0\\
1 & 1 & 1 & 0 & 0\\
0 & 1 & 1 & 1 & 0\\
0 & 0 & 1 & 1 & 1\\
0 & 0 & 0 & 1 & 1\\
\end{pmatrix},
\end{equation}
that is, we assume variables align in a line with 
each one only correlated with its nearest neighbors. 
We generate daily time series for an arbitrary date range
from January 1, 2016 to April 9, 2016,
with a perturbation beginning on March 21, 2016. 
We specify dates in the simulation to facilitate the
description of the intervention period.
We first generate five multivariate datasets for test stores with varying levels of impact and label them as  
Datasets 1--5.

For each Dataset $i$, $i = 1, \dots, 5$, the trend is generated from
$\mu_{it} \sim \mathcal{N}(0.8\mu_{i,t-1}, 0.1^2)$
with $\mu_{i0} = 1$.
The weekly components are generated from two sinusoids of
the same frequency 7 as follows:
\begin{equation}
\label{equ-4.2}
 \delta_{it}
 = 
 0.1 \times \cos({2\pi t}/{7}) + 0.1 \times \sin({2\pi t}/{7}).
\end{equation}
Additional datasets for 10 control stores are generated,
each from an AR(1) process
with coefficient $0.6$ and standard error $1$.
We let the first and second datasets to
have regression coefficients
$\beta_1 = 1$, $\beta_2 = 2$ and 
let the rest to be 0.
We then generate 
the residuals $\boldsymbol \epsilon_t$ 
in the observation equation from the multivariate normal distribution $\mathcal{N}(\boldsymbol 0, \boldsymbol \Sigma)$ with precision matrix having sparsity structure given by (\ref{equ-4.1}).
We set the diagonal elements for $\boldsymbol \Sigma^{-1}$ to 10,
and its non-zero off-diagonal elements to 5.
The simulated data for test stores are the sum of the simulated values of
$\boldsymbol \mu_t$, $\boldsymbol \delta_t$,
$\boldsymbol X_t \boldsymbol \beta$ and $\boldsymbol \epsilon_t$.
The causal impacts are generated as follows:
for each Dataset $i$, $i=1,\ldots,5$, 
we add an impact scale $\frac{(i-1)}{2} \times (\log 1, \dots, \log 20)$ 
from March 21, 2016 to April 9, 2016. Clearly no causal impact is added in Dataset 1. 

We impose the graphical structure with adjacency matrix in (\ref{equ-4.1}) in both observed and hidden processes 
in the model
and then apply the two-stage algorithm to estimate parameters.
In Stage 1, we apply the revised EMVS algorithm.
We choose the initial values $\boldsymbol \beta^{(0)}$ and $\boldsymbol a_1^{*(0)}$ to be the zero vectors
and the first $15 \times 15$ elements of $\boldsymbol P_1^{*(0)}$, 
which correspond to the covariances of the trend, local trend and seasonality components, to be a diagonal matrix. 
The remaining elements in $\boldsymbol P_1^{*(0)}$ are set to 0.
We select 20 equally spaced relatively small values for $v_0$ from 
$10^{-6}$ to $0.02$ and a relatively larger value for $v_1$, $10$.
For the prior of $\theta$, we set $\zeta_1 = \zeta_2 = 1$.
The maximum number of iterations of the EMVS algorithm is chosen to be 50.
We calculate the threshold of non-zero value of $\boldsymbol \beta_i$ from the inequality:
$p(\gamma_i = 1| \beta_i, \boldsymbol Y_t^*, \boldsymbol X_t^*) > 0.5$
\citep[See the detailed discussions in][]{rock14}.
Then the threshold can be expressed as
\begin{equation*}
	|\beta_i^\text{th}| \geq 
	\sqrt{\frac{\log(v_0/v_1) + 2\log(\hat\theta/(1-\hat\theta))}{v_1^{-1} - v_0^{-1}}},
\end{equation*}
where $\hat \theta$ is the maximized value obtained from the EMVS algorithm. 
\citet{rock14} also suggested using a  
deterministic annealing variant of the EMVS (DAEMVS) algorithm
which maximizes
\begin{equation}
\label{equ-4.3}
      \mathbb{E}_{(\boldsymbol \alpha_{1:T}^*, \boldsymbol \gamma)| \cdot} 
       \big[ 
                \frac{1}{s}
                \log \pi(\boldsymbol \alpha_{1:T}^*, \boldsymbol \beta, 
    \boldsymbol \gamma, \theta, \boldsymbol \Phi,
    \boldsymbol \Sigma, \boldsymbol Q
    | \boldsymbol Y_t^*, \boldsymbol X_t^*)^s\ | 
    \ \boldsymbol \beta^{(k)}, \theta^{(k)}, \boldsymbol \Phi^{(k)}, \boldsymbol \Sigma^{(k)}, \boldsymbol Q^{(k)}
       \big],
\end{equation}
where $0 \leq s \leq 1$.
The parameter $1/s$ is known as a temperature function \citep{ueda98}.
When the temperature is higher, that is when $s \rightarrow 0$,
the DAEMVS algorithm has a higher chance to find a global mode
and thus reduces the chance of getting trapped at 
a local maximum. 

\begin{figure}[!h]
\centering
    \makebox[\textwidth][c]{\includegraphics[width=1.1\textwidth]{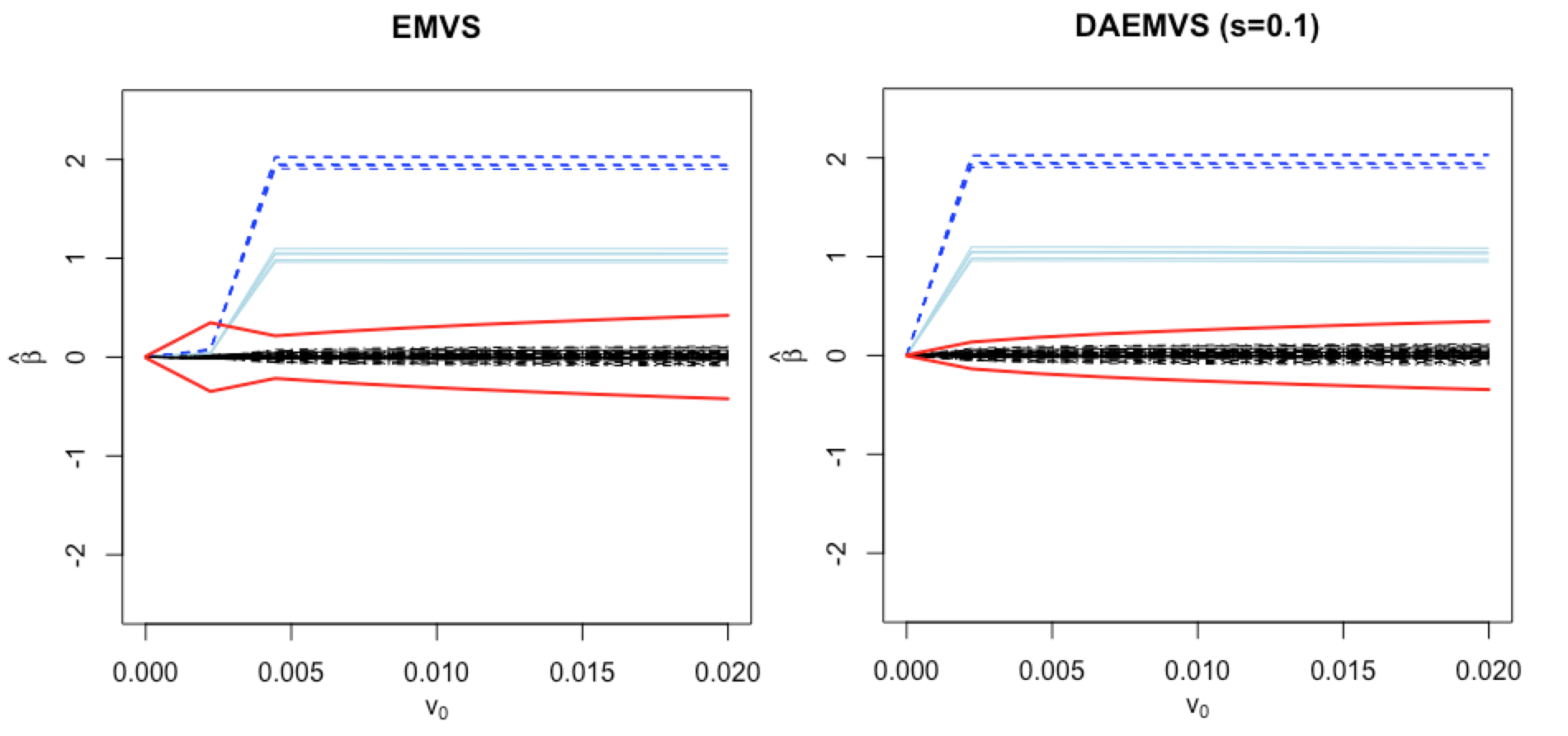}}
   \caption{\small EMVS (left) and DAEMVS (with $s = 0.1)$ (right)
   estimation of $\boldsymbol \beta$ based on the simulated datasets. 
   The dark blue lines are the parameters that have simulated values $2$;
   the light blue lines are the parameters that have simulated values $1$
   and the black lines are the parameters that have simulated values $0$.
   The red lines are the calculated $\beta_i^\text{th}$ values,
   within the two red lines, the parameters should be considered as zero parameters.}
   \label{fig-1}
\end{figure}

Figure \ref{fig-1} compares the results for 
using EMVS and DAEMVS with $s = 0.1$ algorithms.
We plot $\hat{\boldsymbol \beta}$ and their thresholds 
based on 20 different values of $v_0$ from $10^{-6}$ to $0.02$.
From the plot, the estimated values for $\boldsymbol \beta$ using both 
EMVS and DAEMVS methods are close to their true values. 

The true zero coefficients are estimated to be very close to 0.
However,
we observe that the values of $\beta_i^\text{th}$ is larger
by using the EMVS method compared to the DAEMVS method.
This is because in the region where $v_0$ is less than 0.005, 
the $\hat\theta$ estimated from EMVS is very close to 0, thus the negative value of $\log\big(\hat\theta/(1-\hat\theta)\big)$ is very large and the threshold becomes larger.
Based on the simulation results, we use DAEMVS with $s=0.1$ throughout the rest of the paper.

The DAEMVS gives a smaller value of $\beta_i^\text{th}$, 
yet the thresholds can distinguish the true zero and non-zero coefficients in this case.  
Nevertheless it may miss a non-zero coefficient if the coefficent is within the thresholds.
In practice,
since our goal is to identify significant control variables
and use them to build counterfactuals for a causal inference,
we may choose to include more variables than the threshold suggests provided that the total number of included variables is still manageable.

\begin{figure}[!h]
\centering
    \makebox[\textwidth][c]{\includegraphics[width=1.1\textwidth]{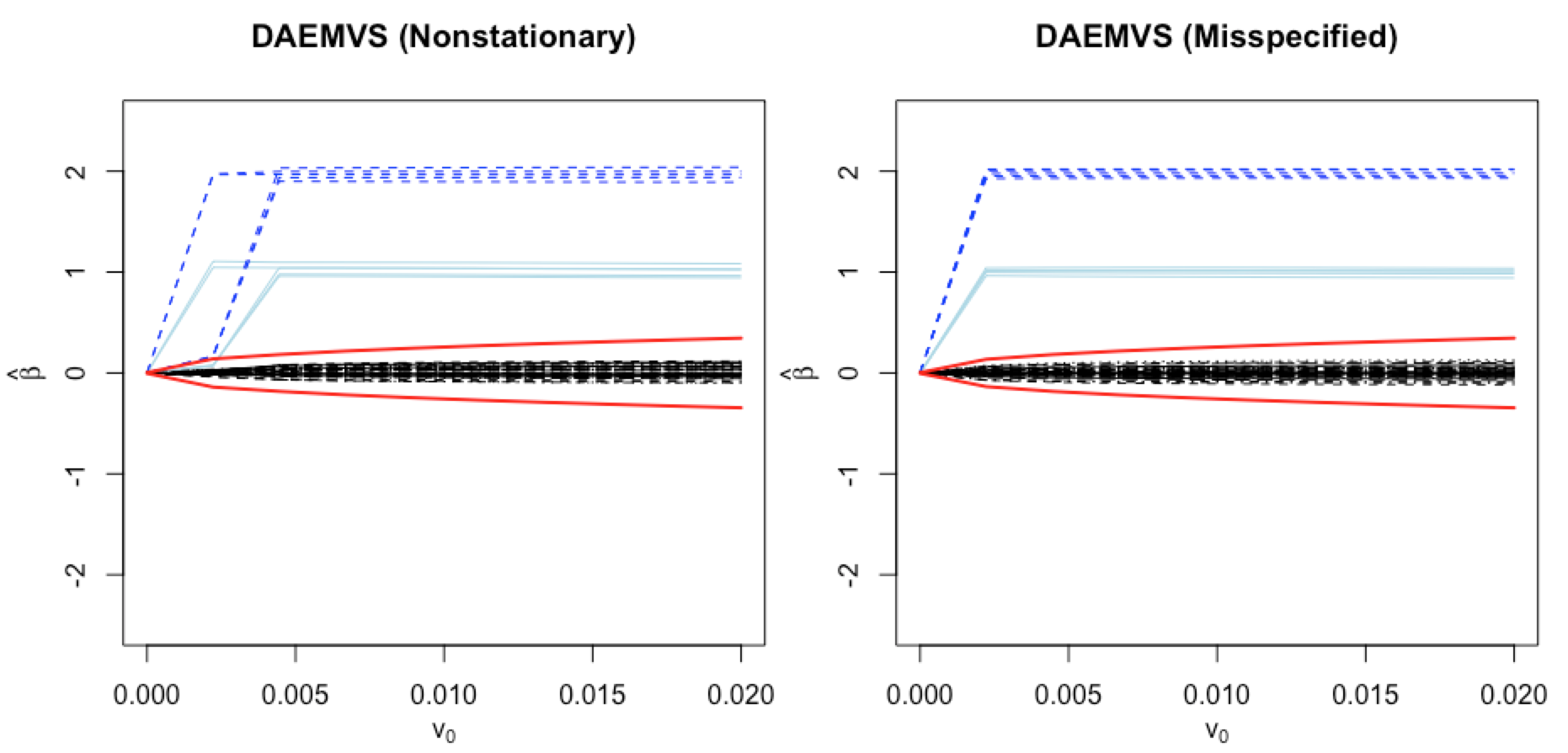}}
   \caption{\small DAEMVS (with $s = 0.1)$ 
   estimation of $\boldsymbol \beta$ based on the simulated datasets using the nonstationary model (left) and the misspecified model (right). 
   The dark blue lines are the parameters that have simulated values $2$;
   the light blue lines are the parameters that have simulated values $1$
   and the black lines are the parameters that have simulated values $0$.
   The red lines are the calculated $\beta_i^\text{th}$ values,
   within the two red lines, the parameters should be considered as zero.}
   \label{fig-1-nm}
\end{figure}

Recall that in the Stage 1, we used a conjugate prior for $\text{vec}(\boldsymbol \Phi)$ instead of the originally proposed prior described in Section \ref{sec-3.3}. 
Here, we want to make sure the change of prior would not affect the results of $\hat{\boldsymbol \beta}$ too much.
We conduct the analysis by choosing two different values of the covariance matrix of the prior: $\boldsymbol I_5$ and $0.01 \times \boldsymbol I_5$.
We found the estimates $\hat{\boldsymbol \beta}$s are almost identical to the estimated values shown in Figure \ref{fig-1}.
We also consider using other two models: one ignores the stationarity constraint for $\boldsymbol \tau_t$ (henceforth the ``nonstationary model'');
another ignores the time dependency of the model (henceforth the ``misspecified model'').
To be more explicit,
for the nonstaionary model, we let the local linear trend follow
(\ref{equ-2.4}). 
The misspecified model is given by $\boldsymbol Y_t^* = \boldsymbol X_t^* \boldsymbol \beta + \boldsymbol \varsigma_t$, with $\boldsymbol \varsigma_t$s are i.i.d random errors with multivariate normally distributed and mean $\boldsymbol 0$ by ignoring their dependency.
We conduct DAEMVS with $s=0.1$ for both of the two models.
In the nonstationary model, we choose a diffuse prior for $\boldsymbol \alpha^*_1$ and change the covariance corresponding to the local linear trend in $\boldsymbol P_1^{*(0)}$ to be $10^6 \times \boldsymbol I_5$.
In the misspecifed model, the M-step can be simplified to only updates for $\boldsymbol \beta$, $\theta$ and the covariance matrix of $\boldsymbol \varsigma_t$.
We plot the results into Figure \ref{fig-1-nm}.
Comparing the results in Figure \ref{fig-1-nm} with Figure \ref{fig-1}, there are not much differences among the results obtained using the three different models for estimating $\boldsymbol \beta$.

In Stage 2, we plug-in $\hat{\boldsymbol \beta}$ and calculate $\widetilde{\boldsymbol Y}_t$ in (\ref{equ-2.6new}).
We choose the prior for the rest of parameters as follows:
we let $\boldsymbol \alpha_1 
\sim 
\mathcal{N}(\boldsymbol 0, \boldsymbol I)$.
If $\boldsymbol \tau_t$ is a nonstationarity process, 
the initial condition is considered as a diffuse random variable with large variance \citep{durb02}.
Then we let
the covariance matrix of $\boldsymbol \tau_t$ to be $10^6 \times \boldsymbol I_5$.
We let $\nu = 1$, $k_1 = k_2  = k_3 = 0.1$.
We choose $\boldsymbol H = \boldsymbol I_5$ and 
the priors for $25$ parameters decomposed 
from $\boldsymbol \Phi$ to be $\mathcal{N}(0, \sqrt{5}^2)$,
and let $\iota \sim \text{Bernoulli}(0.5)$.
We run total $10,000$ MCMC iterations with 
the first $2,000$ draws as burn-in.
An MCMC convergence diagnostic and a sensitive analysis of the model
are conducted, we include their results in the supplementary file.

\subsection{Performance of the commonly used causal inference method}
\label{sec-5.2}

\begin{figure}[!h]
\centering
    \makebox[\textwidth][c]{\includegraphics[width=1.1\textwidth]{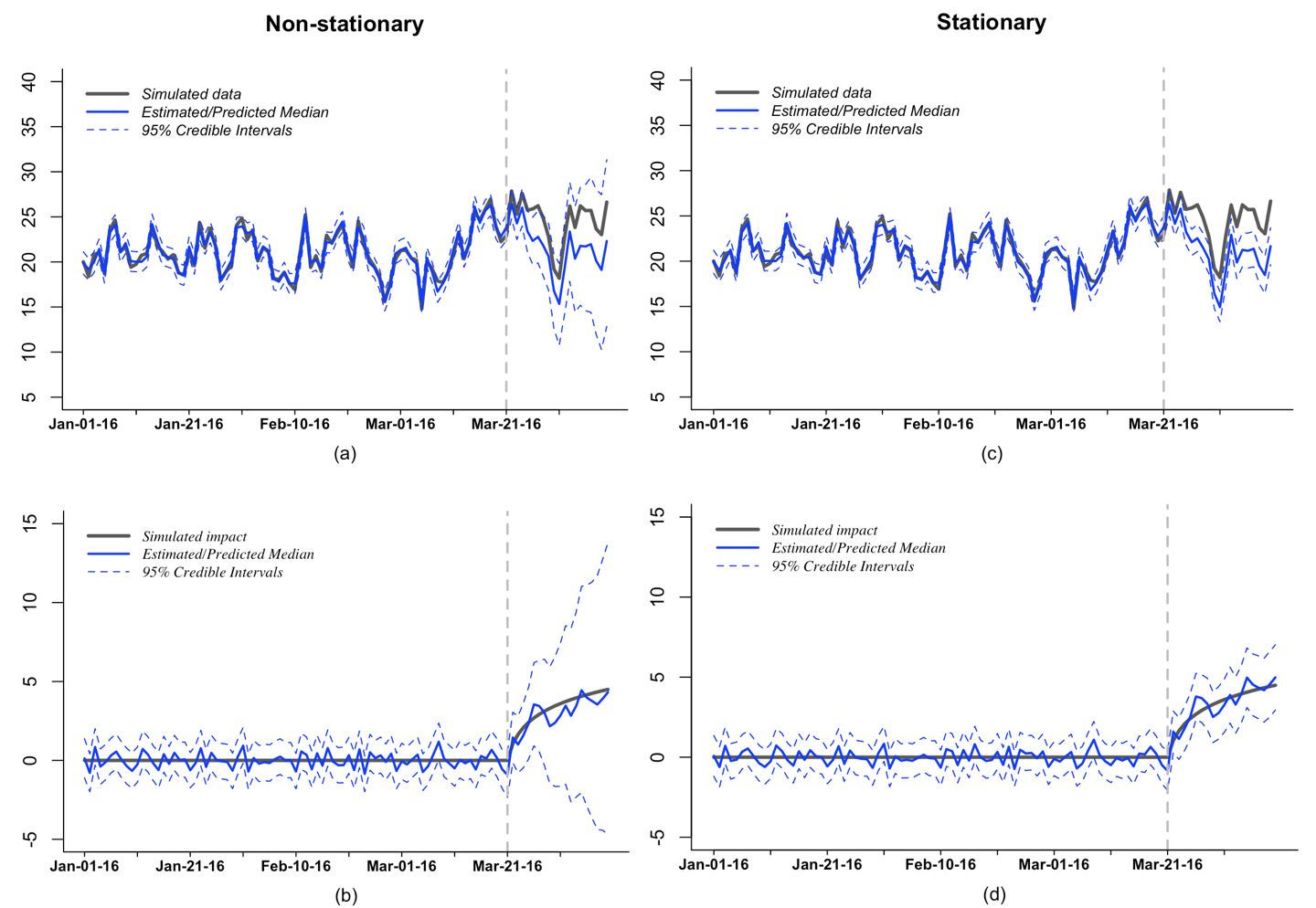}}
   \caption{\small Plot of the causal impact in Dataset $4$ using models with a stationary and a nonstationary local linear trend.    
   (a) and (c) are the plots of estimation (before March 21, 2016) and 
   prediction (after March 21, 2016)
   of Dataset 4 without stationarity constraint (left) 
   and with stationarity constraint (right). 
   The gray line is the simulated dataset, the blue line is the estimated 
   posterior median of the dataset using the model,
   the dashed blue line is the corresponding $95\%$ credible and prediction 
   intervals.
   (b) and (d) are the plots of estimated causal impact by 
   taking the difference between the observed data
   and Bayesian estimates using the model with a nonstationary local linear trend (left) 
   and the model with a stationary local linear trend (right). 
   The black line is the simulated true impact, 
   the blue line is the estimated median of the impact,
   the dashed blue lines are the corresponding $95\%$ credible and prediction intervals.}
   \label{fig-2}
\end{figure}

In this section, we study the performance of the commonly used method.
The causal effect is estimated by taking the difference between
observed data during causal period and the potential outcomes of counterfactuals during that period.
In Stage 1, 
we use the DAEMVS ($s = 0.1$) algorithm to estimate $\hat{\boldsymbol \beta}$
for the model (\ref{equ-2.6})--(\ref{equ-2.7}). 
A stationarity constraint is added on the local linear trend $\boldsymbol \tau_t$.
In Stage 2, 
we consider two different settings for $\boldsymbol \tau_t$:
with and without adding the stationarity constraint.
We choose Dataset 4 as an example and plot accuracy of the model based on the two different settings in Figure \ref{fig-2}.
There are four subplots:
the left two subplots are the results for the model with a nonstationary local linear trend and the right two subplots are the results for the model with a stationary local linear trend.
Before the period with a causal impact, which is March, 21, 2016, 
the estimated posterior medians and 95\% credible intervals obtained from the two models are close (see plots (b) and (d) in Figure \ref{fig-2});
but their prediction intervals during the period with a causal impact are quite different. 
In the model with a nonstationary local linear trend, the prediction intervals 
are much wider and expand more rapidly 
than those resulting from the model 
with a stationary local linear trend. 
In the former case, the observed data during the campaign are fully contained inside 
the prediction intervals and thus failed to detect a causal impact.
However, the model with a stationary local linear trend gives only moderately increasing 
prediction intervals and thus can detect the causal impact. 
Plots (b) and (d) shown in the bottom of Figure \ref{fig-2} are the
estimated causal impact in each model for Dataset 4
calculated by taking the difference between 
observed values and counterfacutal potential outcomes. 
In each plot, the estimated causal impact 
medians are able to capture the shape of the simulated causal impact. 
However, the prediction intervals in plot (b) contain the value 0 and thus negate the impact. The shorter prediction intervals in plot (d) do not contain the value 0,
and thus indicate the existence of a impact.

To give an overall picture of the model fitting for the five simulated datasets, 
we summarize the posterior medians and their $95\%$ credible intervals of the estimated causal impact for 
all the datasets in Table \ref{tab-2}.
In the model with a nonstationary local linear trend, 
no impacts are detected for all the five datasets 
since their corresponding prediction intervals all contain the value 0. 
In the model with a stationary local linear trend on $\boldsymbol \tau_t$, 
the impacts are successfully detected for the last three datasets. 
For Dataset 2, it has a weaker impact.
Its impact is not detected even after imposing the stationarity constraint.
Also, when the stationarity constraint is imposed, including 
the intercept $\boldsymbol D$ in (\ref{equ-2.3}) helps
give a robust long run prediction. 
Thus, from Table \ref{tab-2}, we find that the estimated medians 
using the model with a stationary local linear trend are closer to the true impact 
compared with that obtained from using the model with a nonstationary local linear trend.

In the setting where the sales in the test stores are spatially correlated, the use of the multivariate model with a stationary local linear trend is necessary for obtaining more accurate estimates for causal effects. 
We compare the results with a univariate model which ignores the correlation between the five simulated datasets.
We fit the five datasets independently into that model.
The model is the univariate version of the model (\ref{equ-2.6})--(\ref{equ-2.7}). 
In the univariate model, 
the errors $\boldsymbol \epsilon_t$, $\boldsymbol u_t$, $\boldsymbol v_t$ and $\boldsymbol w_t$
become scalars.
We denote $\sigma^2$, $\sigma_u^2$, $\sigma_v^2$ and $\sigma_w^2$ as their corresponding variances. 
We choose their priors as
$\sigma^{-2} \sim \text{Gamma}(\frac{0.1}{2}, \frac{0.1\times \text{SS}}{2})$,
$\sigma_u^{-2}, \sigma_v^{-2}, \sigma_w^{-2} 
\sim 
\text{Gamma}(0.01, 0.01\times \text{SS})$,
where $\text{SS} = \sum_{t=1}^T (y_t - \bar{y})^2 / (T-1)$ and
$\bar{y} = \sum_{t=1}^T y_t / T$.
The parameters $\boldsymbol D$ and $\boldsymbol \Phi$ in (\ref{equ-2.3}) also become scalars and to be denoted by $d$ and $\phi$ respectively.
We give them the priors
$d \sim \mathcal{N}(0, 0.1^2)$
and $\phi \sim \mathcal{N}(0, 0.1^2) \mathbbm{1}_{(-1, 1)}$.


\begin{table*}
\small
\centering
\caption{\small Posterior medians and 95\% credible intervals of average causal impacts
              for simulated datasets estimated using the multivariate models with
              a stationary and a nonstationary
              local linear trend.}
\label{tab-2}
\begin{tabular}{cccc}
\toprule
& Simulated impact & Nonstationary & Stationary \\
\midrule
Dataset 1 & 0.00 & 0.00  $[-4.419, 4.425]$ & 0.29  $[-1.440, 1.996]$  \\
Dataset 2 & 1.06 & 0.64  $[-3.989, 5.298]$ & 1.07 $[-0.648, 2.780]$ \\
Dataset 3 & 2.12 & 1.22  $[-3.758, 5.965]$ & 2.27  $[0.399, 4.014]$ \\
Dataset 4 & 3.18 & 2.83  $[-1.793, 7.575]$ & 3.16  $[1.500, 4.862]$ \\
Dataset 5 & 4.23 & 4.25 $[-0.249, 8.771]$ & 4.25  $[2.520, 5.904]$
\\
\bottomrule
\end{tabular}
\end{table*}

In order to make the comparison between the multivariate model and the univariate model meaningful, we plug-in the same $\hat{\boldsymbol\beta}$ obtained from Stage 1 for both models. 
We conduct an MCMC alogrihtm for the five datasets separately using the univariate model
by sequentially sampling draws from the corresponding posterior distributions of $\alpha_{1:T}$, $d$, $\phi$, $\sigma^2$, $\sigma_u^2, \sigma_v^2$ and $\sigma_w^2$.
We run the MCMC algorithm for 10,000 iterations and treat the first 2,000 as burn-in.
The estimated causal impacts are shown in Table \ref{tab-2-univ}. 
By comparing the results with the results in Table \ref{tab-2},
the univariate model produces wider credible intervals for all of datasets even though their posterior medians are close to the truth. 
Thus the multivariate model with a stationary local linear trend is more accurate for detecting a causal impact.

\begin{table*}
\small
\centering
\caption{\small Posterior medians and 95\% credible intervals of average causal impacts for simulated datasets estimated using the univariate model.}
\label{tab-2-univ}
\begin{tabular}{cccc}
\toprule
& Simulated impact & Stationary (univariate) \\
\midrule
Dataset 1 & 0.00 & 0.17  $[-2.197, 2.472]$  \\
Dataset 2 & 1.06 & 1.03  $[-1.365, 3.473]$ \\
Dataset 3 & 2.12 & 2.16  $[-0.370, 4.476]$ \\
Dataset 4 & 3.18 & 3.20  $[0.821, 5.748]$ \\
Dataset 5 & 4.23 & 4.08  $[1.564, 6.489]$
\\
\bottomrule
\end{tabular}
\end{table*}

We conduct additional independent 10 simulation studies by generating datasets using the same scheme which described above, but using different random number generators from the software. We conduct the same analysis for the 10 simulated studies using the multivariate model with stationarity constraints. 
All of these studies show that
the commonly used method failed to detect causal effect for the second dataset, which is the one with the smallest amount of simulated causal impact.

%

\subsection{Performance of the new method to infer causality}
\label{sec-4.2}

In this section, we study the performance of the new method.
We use the same simulated data in Section \ref{sec-4.1}.
We calculate the one-sided KS distance 
in (\ref{equ-4.1.2})
and the threshold in (\ref{equ-4.1.3})
for each $i = 1, \dots, n$.
We also calculate the one-sided KS distances
\begin{equation*}
\begin{split}
	& \sup_x \Big[  \frac{1}{k} \sum_{j=1}^{k} \big(
	\mathcal{F}(\sum_{t=T+1}^{T+m}
	\mu_{it} \leq x| 
	\boldsymbol Y_{1:T}^{\text{obs}},
	\boldsymbol Y_{T+1:T+m}^{\text{cf}(j)},
	\boldsymbol X_{1:T+m}) \big)\\
	& \quad \quad \quad \quad - 
	\mathcal{F}(\sum_{t=T+1}^{T+m} 
	\mu_{it} \leq x| 
	\boldsymbol Y_{1:T+m}^{\text{obs}},
	\boldsymbol X_{1:T+m})
	\Big]
\end{split}	
\end{equation*}
and the corresponding thresholds for $m = 1$ to $m = P$.
This allows to see how the KS distances grow over time.

\begin{figure}
\centering
    \makebox[\textwidth][c]{\includegraphics[width=1.1\textwidth]{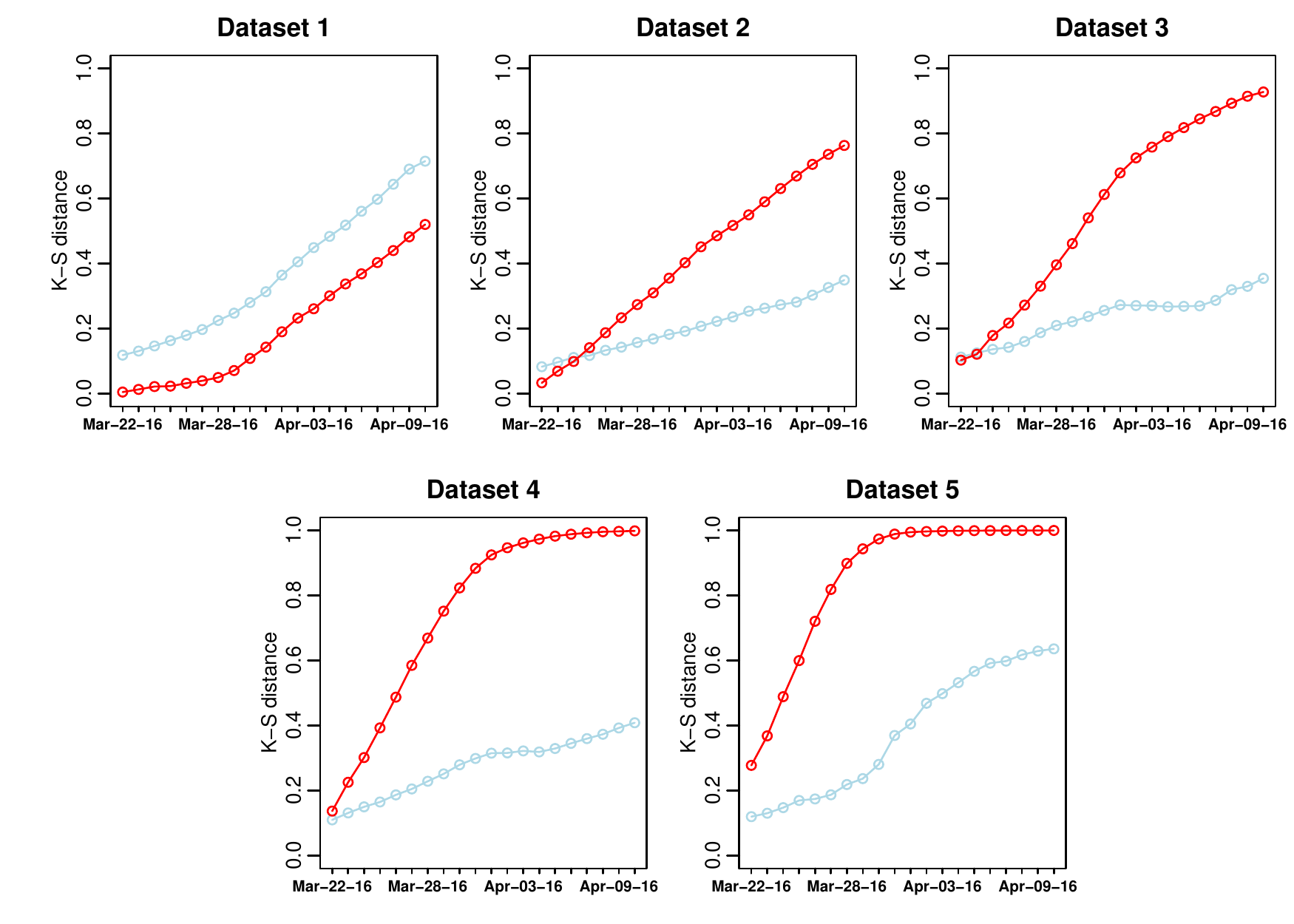}}
   \caption{\small 
   Results of applying the new method to detect causal impacts in Dataset 1, $\dots$, Dataset 5 using the multivariate model with a stationary local linear trend during the causal period from March, 22, 2016 to April, 9, 2016.
   In each subplot, the red line gives the one-sided KS distances between two posterior distributions with one is given the data of counterfactuals; the light blue line gives the corresponding thresholds.}
   \label{fig-5}
\end{figure}

We plot the results in Figure \ref{fig-5}.
There are five subplots in that figure with each represents one simulated dataset.
For each subplot, 
the red line represents the one-sided KS distances between posteriors from a test store and its counterfactuals,
and the lightblue line represents its corresponding thresholds.
The threshold is calculated based on 
$k = 30$ simulated counterfactual datasets.
In the plot,
Dataset 1 is the only one with the one-sided KS distances completely below the thresholds
and it is the dataset which does not receive any impacts.
This suggests that our method has successfully distinguished between impact
and no impact in these datasets. 
For Dataset 2, the impact at the early period is small,
thus we observe the causal impact in the first three predicting periods are not significant;
however, the new method can detect the impact after the fourth period. 

We also summarized the results in Table \ref{tab-ks-multi}.
Compared with the results from the commonly used method (see Table \ref{tab-2}), 
the new method shows a significant improvement in detecting causal impacts.
From Dataset 3 to Dataset 5, the one-sided KS distances are all above their corresponding thresholds.
Also, as the impact grows stronger, 
we observe that the distances becomes larger.
The thresholds too increase along the time,
since the predicting intervals for the trends become wider.

To check the performance of the new method,
we conduct 10 more simulation studies using the data generated from the same model.
Although the values of the one-sided KS distances and thresholds are not identical for each simulation, since the model is highly flexible and the estimated trend is sensitive to local changes of a dataset,
the new method successfully detects the causal impacts in Dataset 2, \dots, Dataset 5.

\begin{table*}
\small
\centering
\caption{\small Results of the one-sided KS distances and thresholds obtained by
applying the new method to detect causal impacts in Dataset 1, \dots, Dataset 5 using the multivariate model with a stationary local linear trend.
We only present the results at the dates March 22, 2016, March 31, 2016 and April. 9, 2016 which correspond to the 1st day, 10th day and 20th day during the causal period.
}
\label{tab-ks-multi}
\begin{tabular}{ccccccc}
\toprule
& & Dataset 1 & Dataset 2 & Dataset 3 & Dataset 4 & Dataset 5\\
\midrule
March, 22 & KS distance & 0.005 & 0.033 & 0.103 & 0.137 & 0.277 \\
(1st day)& Threshold & 0.118 & 0.083 & 0.112 & 0.110 & 0.120 \\
\midrule
March, 31 & KS distance & 0.143 & 0.402 & 0.612 & 0.884 &	0.989 \\
(10th day)& Threshold & 0.313 & 0.192 & 0.256 & 0.299 & 0.369 \\
\midrule
April, 9 & KS distance & 0.520 & 0.763 & 0.928 & 0.999 &	1.000\\
(20th day)& Threshold & 0.715	& 0.349 & 0.354 & 0.409 & 0.636 \\
\bottomrule
\end{tabular}
\end{table*}

We applied the new method to the univariate model,
which is described in Section \ref{sec-5.2},
using the same simulated dataset.
The graphical and tabular representations of the results are presented in Section 3 of 
the supplementary material.
We found that by comparing 
with the results obtained from the multivariate model (see Figure \ref{fig-5}), the thresholds are much larger among all the datasets. Recall that from Table \ref{tab-2-univ}, the credible intervals estimated using the commonly used method are wider. Thus when we randomly draw samples from a counterfactual with a larger variance, the posterior distributions for their trend are more apart. As a result, the pairwise one-sided KS distances between the posterior distributions of the trends are larger.  
Even though the thresholds are larger when using the univariate model, unlike the results obtained by using the commonly used method, the new method can still detect the causal impact for almost all the datasets which received an impact successfully, except for the very weak impacts in Dataset 2 during the first three periods and Dataset 3 during the first period.


\section{Application to a real dataset}
\label{sec-6}

In this section, we present the results of a real data analysis for measuring the 
causal impact of an online advertising campaign (run by {\it MaxPoint}) 
for a consumer product at a large national retail chain. 

Due to commercial confidentiality, we do not show full details of the results,
but the following description explains how our method works
in this real dataset.
{\it MaxPoint} targets this campaign at 627 test stores
and 318 control stores
spread out across the country and collects weekly data throughout the campaign.
We choose all the control stores in the corresponding state for each dataset. 
If a state does not at all have control stores, 
we remove such data from the analysis.
In Stage 1, we use the DAEMVS (with $s = 0.1$) algorithm to select the control stores
for each test store. 
If for a test store, all the potential control stores are eliminated by the DAEMVS algorithm
we also eliminate that store from the causal analysis, 
because without building a counterfactual, 
the causal inference cannot be conducted. 
After making the selection, we conduct the causal analysis on 323 test stores in total.
For each dataset, there are 46 weekly observations in total with the last 10 
observations occurring in the causal period. 
Since the length before the causal period is only 35 per dataset,
we have to separate these 323 stores into smaller datasets 
and fit the model separately on them.
As large national chain retailers organize promotional and operations activity
differently in each state, 
we treat stores in different states as independent. 
State-wise splitting typically keeps the number of stores less than 15.
If one state has more than 15 stores,
we split further into subregions to meet the requirement. 
We further assume that the stores in two different subregions behave independently.
The regions are separated based on city boundaries.
Within each region,
we assume that stores are connected with each other.
This means that the inverse covariance matrix (equivalently, the covariance matrix) follows a block-diagonal structure
with at most 15 nodes in a block. 

We assume the three causal assumptions in Section \ref{sec-1} hold.
The following table summarizes the number of stores with significant causal effects
from the advertising campaign.
From the table we found that
the number of stores are increasing from the first week to the last week. 
During the first five weeks, the number of stores that received 
causal impact increased rapidly compared with that in the last five weeks.

\begin{table*}[!h]
\small
\centering
\caption{\small Number of test stores that received significant causal impacts for each week
of running the advertisement campaign by using the multivariate model with a stationary local linear trend.}
\label{tab-3}
\begin{tabular}{cccccc}
\toprule
 & 1st week  & 2nd week & 3rd week  &
				4th week & 5th week\\
Number of stores & 23 & 44 & 55 & 62 & 73\\
\midrule
& 6th week  & 7th week & 8th week  &
				9th week & 10th week\\
Number of stores & 72 & 77 & 78 & 82 & 84
\\
\bottomrule
\end{tabular}
\end{table*}

\begin{figure}[!h]
\centering
\begin{minipage}{.5\linewidth}
\centering
\subfloat[2nd week]{\label{fig-9a}\includegraphics[scale=.43]{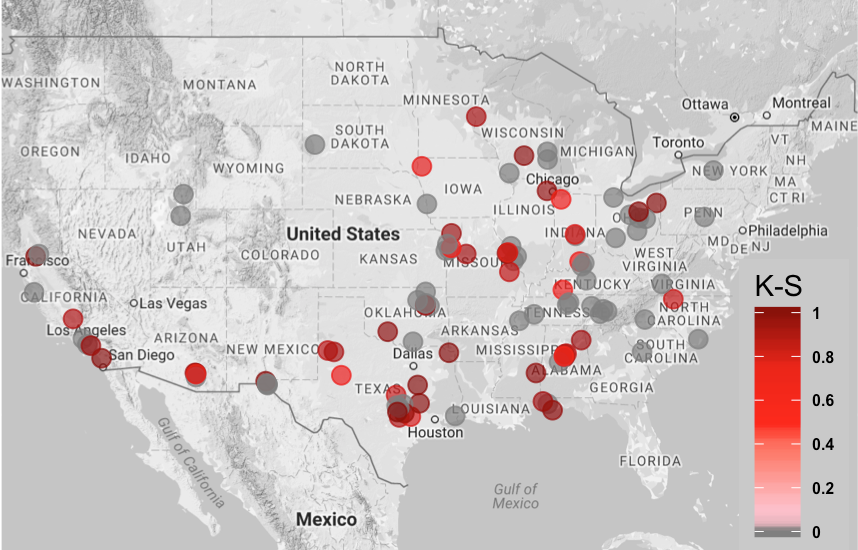}}
\end{minipage}%
\begin{minipage}{.5\linewidth}
\centering
\subfloat[5th week]{\label{fig-9b}\includegraphics[scale=.43]{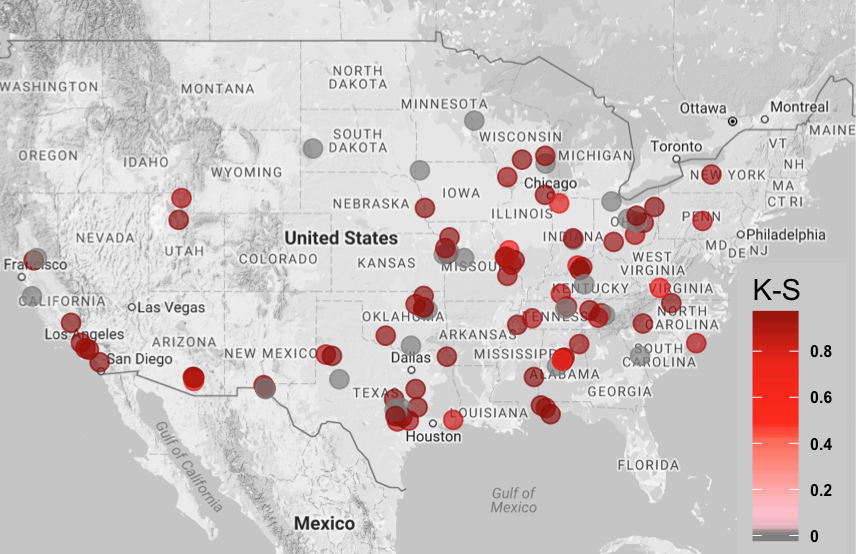}}
\end{minipage}\par\medskip
\centering
\subfloat[10th week]{\label{fig-9c}\includegraphics[scale=.43]{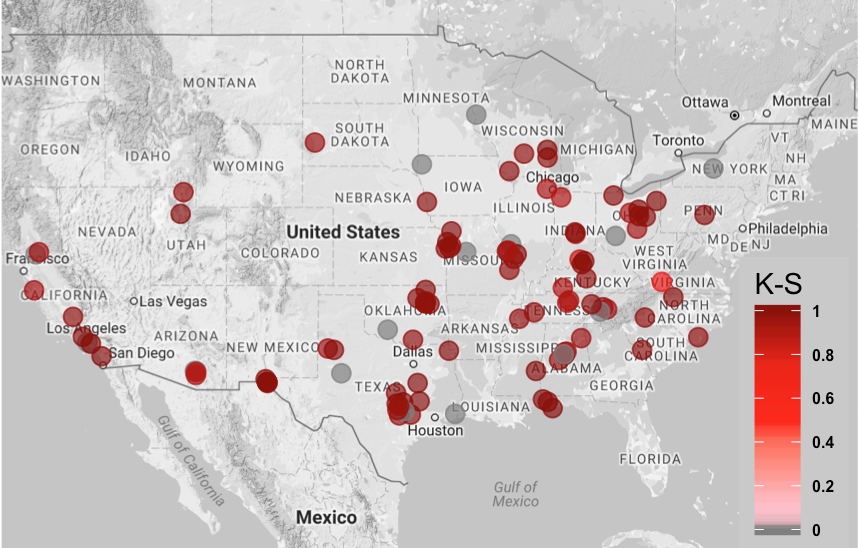}}
\caption{
Plot of the causal impacts at test stores at end of the second week (a), the fifth week (b)
and the last week (c) for an advertising campaign of a consumer product at a large national retail chain. 
The impacts below their thresholds are set to zero.
The United States map is produced using Google Maps, 2017.}
\label{fig-9}
\end{figure}

Not only the number of impacted stores increased during the
advertising campaign period (shown in Table \ref{tab-3}), 
the magnitudes of the impacts in those stores also increased. 
In Figure \ref{fig-9},
we plot the estimated one-sided KS distances for stores 
along with their locations at Weeks 2, 5 and 10.
In each figure, we only plot the stores with significant causal effects.
The red dots represent the stores with the one-sided KS distances larger than their 
corresponding thresholds, which suggests that those stores received significant causal effects. 
The grey dots represent the stores that do not show significant causal effects.
We find that the magnitudes of the impacts for most of the stores have a larger
increase from the first five weeks compared with the last five weeks.
Comparing the plots of the fifth week and the tenth week,
we find that only a few stores in California, South Dakota, Ohio and Texas got increased causal effects. 

We also conduct an analysis by assuming that the test stores are independent and thus ignoring their spatial correlation. Table \ref{tab-3-u} lists the number of stores that received significant causal effects. The numbers are smaller than those obtained using the multivariate model.
This suggests most of the impacts are weak and the spatial correlation between sales in different stores help detect the weaker impacts.

\begin{table*}[!h]
\small
\centering
\caption{\small 
Number of test stores that received significant causal impacts for each week
of running the advertisement campaign by 
using the univariate model.}
\label{tab-3-u}
\begin{tabular}{cccccc}
\toprule
 & 1st week  & 2nd week & 3rd week  &
				4th week & 5th week\\
Number of stores & 25 & 19 & 22 & 23 & 18\\
\midrule
& 6th week  & 7th week & 8th week  &
				9th week & 10th week\\
Number of stores & 17 & 15 & 15 & 13 & 14
\\
\bottomrule
\end{tabular}
\end{table*}

\section{Conclusion and discussion}
\label{sec-7}
In this paper, we proposed a novel causal inference method
which compares the posterior distributions of the latent trend conditional on two 
different sets of data: 
one is the observed data which contain a causal effect;
the other one is the data from a synthetic control. 
We calculated the one-sided KS test statistics between the two posterior distributions.
A threshold was used to decide whether a causal impact is significant or not.
In the simulation study, we showed that our method can detect a smaller sized causal impact more efficiently
compared with the commonly used method even when the model is slightly misspecified.
The new causal inference method is not restricted to the specific structural time series model used in this paper and can be applied to many other models in different applications.

We used a multivariate structural time series model to 
estimate the causal impact of a stimulus on subjects such as an advertising campaign for each individual store. 
Sales in those stores are spatially correlated. 
A Bayesian analysis was used to estimate parameters in this model. 
We imposed sparsity on the precision matrix based on the distance between each pair of stores. The sparsity was imposed through a $\mathcal{G}$-Wishart prior, where the graph $\mathcal{G}$ can be either decomposable or non-decomposable. 
We restricted the hidden process $\boldsymbol \tau_t$ to be stationary in order to stabilize the prediction intervals. 
To sample its time-varying variables, we used the Kalman filter and simulation smoother algorithm.
This algorithm can be used to impute missing values inside the MCMC loops.

We used the revised EMVS algorithm to select control stores.
We also discussed the advantage of using the DAEMVS algorithm which is a modified version of the EMVS algorithm.
Compared to the EMVS algorithm,
the DAEMVS algorithm reduces the chance of getting trapped at a local maximum.
Both the EMVS and DAEMVS algorithms are computationally much faster than the sampling based method like SSVS. 
Since the EMVS algorithms cannot be incorporated into MCMC loops, we proposed a two-stage algorithm to estimate parameters. In Stage 1, we used the DAEMVS to obtain $\hat{\boldsymbol \beta}$; in Stage 2, we plugged-in $\hat{\boldsymbol \beta}$ and used an MCMC algorithm to obtain posterior distributions of the remaining parameters. 

We compared the multivariate model with the univariate model which assumes independence between responses based on simulated datasets. The results indicate that the univariate model gives wider credible intervals (if using the commonly used method) and larger threshold (if using the new method) than the multivariate model. Thus incorporating of the spatial relationships between test stores is beneficial.

Finally, we analyzed a real dataset on sales data of products distributed through 
brick and mortar retail stores for
an advertising campaign run by {\it MaxPoint}. Even though, due to commercial confidentiality, we did not provide the full details of the results, the summarization tables of the number of stores that received significant impact suggests the effectiveness of using the new causal inference method.







\bigskip
\begin{center}
{\large\bf ACKNOWLEDGEMENTS}
\end{center}

We thank {\it MaxPoint Interactive Inc.} for providing funding and datasets for this research. We thank Dr. Alice Broadhead, Mark Lowe, Professor Fan Li, Professor Mike West and the referees for valuable input and suggestions.
We thank Professor Anindya Roy for providing the \textsf{R} code for the stationarity constraint and many helpful discussions. Research of the second author is partially supported by NSF grant number DMS-1510238.

\bigskip
\begin{center}
{\large\bf SUPPLEMENTARY MATERIALS}
\end{center}

Two supplementary materials are provided.
The first one contains five sections.
Sections 1 and 2 provide the details on deriving the two-stage algorithm and the revised EMVS algorithm.
Section 3 provides graphical and tabular representations of the results of the new method to infer causality using the univariate model. 
Section 4 provides model checking results.
Section 5 describes the Kalman filter and backward smoothing algorithm.

The second material includes the original Bayesian multivariate time series model code written in 
\textsf{R}.
The code is also available on the website: 
\url{https://github.com/Bo-Ning/Bayesian-multivariate-time-series-causal-inference}.

\bibliographystyle{chicago}
\bibliography{citation}
\end{document}